\documentclass[nofootinbib,reprint,superscriptaddress,preprintnumbers,longbibliography]{revtex4-2}

\usepackage[bottom]{footmisc}
\usepackage{amsmath,amssymb,marginnote}
\usepackage{multirow} 
\usepackage{booktabs} 
\usepackage{graphicx,subfigure}
\usepackage{dcolumn}
\usepackage{bm}
\usepackage[mathlines]{lineno}
\usepackage[colorlinks=true, allcolors=blue]{hyperref}
\usepackage[usenames]{color}
\usepackage{xcolor}
\usepackage{comment}
\usepackage{physics}
\usepackage{ulem}
\usepackage{algpseudocode}

\usepackage{float}
\makeatletter
\let\newfloat\newfloat@ltx
\makeatother
\usepackage{algorithm}
\makeatletter
\let\newfloat\undefined
\renewcommand{\fnum@algorithm}{Algorithm~\thealgorithm}
\makeatother

\newcommand{\pol}{\bm{P}}
\newcommand{\polscalar}{P}
\newcommand{\mo}{\bm{q}}
\newcommand{\vel}{\hat{\mo}}
\newcommand{\num}{\bm{n}}
\newcommand{\numm}{\bm{m}}
\newcommand{\nummm}{\bm{l}}
\newcommand{\nums}{n}  
\newcommand{\numz}{\bm{0}}
\newcommand{\pos}{\bm{r}}
\newcommand{\wav}{\bm{k}_{\num}}
\newcommand{\wf}{\Psi}
\newcommand{\ham}{\mathcal{H}}
\newcommand{\nl}{N}
\newcommand{\vecs}{S}
\newcommand{\vals}{\Omega}
\newcommand{\co}{\Phi}
\newcommand{\rvec}{\psi}
\newcommand{\lvec}{\hat{\psi}}
\newcommand{\ind}{i}
\newcommand{\indd}{j}
\newcommand{\conj}{*}
\newcommand{\magni}{a}
\newcommand{\phase}{\theta}
\newcommand{\thresh}{C}

\begin{document}

\preprint{LA-UR-26-26490}

\title{Neutrino flavor-wave transport: Numerical tests and theoretical challenges}

\author{Anson Kost}
\affiliation{Department of Physics and Astronomy, University of New Mexico, Albuquerque, NM 87131, USA}

\author{Lucas Johns}
\email{ljohns@lanl.gov}
\affiliation{Theoretical Division, Los Alamos National Laboratory, Los Alamos, NM 87545, USA}

\author{Huaiyu Duan}
\affiliation{Department of Physics and Astronomy, University of New Mexico, Albuquerque, NM 87131, USA}

\begin{abstract}
Neutrino quantum kinetics is computationally intractable in the neutrino-dense arenas of core-collapse supernovae and neutron star mergers. Flavor-wave (or flavomon) transport is an emerging approach to this problem in which small-scale flavor inhomogeneities are treated as quasiparticles with properties determined by the local mean background.  We present the first numerical calculations of flavor-wave transport under slow driving. Our results show some quantitative successes but also underscore the theoretical challenges that need to be overcome. Two issues are particularly concerning. (1) Nonlinear wave--wave coupling may be important. (2) Degeneracies and small energy gaps are responsible for significant nonadiabatic transitions. Future work will need to address these points if flavor-wave transport is to be viable in settings of astrophysical interest. Other topics examined include flavor-wave parallel transport and flavor-space symmetry breaking.
\end{abstract}

\maketitle

\section{Introduction}

Predicting the consequences of neutrino flavor mixing in core-collapse supernovae and neutron star mergers is one of the pressing open questions in theoretical particle and nuclear astrophysics \cite{volpe2024neutrinos, johns2025neutrino, tamborra2025neutrinos}. The overall message of the substantial literature on this topic is that the astrophysical effects are potentially significant. This understanding has been assembled from studies of collective flavor instabilities of three types: fast \cite{sawyer2016neutrino, wu2015effects, wu2017fast, 
wu2017imprints, abbar2019occurrence, delfan2019linear, morinaga2020fast, xiong2020potential, ko2020neutrino, george2020fast, abbar2021characteristics,johns2021fast, nagakura2021where, li2021, just2022, fernandez2022, fujimoto2023explosive, xiong2023collisional, ehring2023, ehring2023fast, ehring2024gravitational, akaho2024collisional, liu2024muon, mukhopadhyay2024time, qiu2025neutrino, wang2025, lund2025angle, mori2025}, slow \cite{kostelecky1993neutrino, duan2006collective, dasgupta2015temporal, shalgar2024neutrino, fiorillo2025theoryslow, fiorillo2025theoryslow2, fiorillo2025first}, and collisional \cite{johns2021collisional, johns2022collisional, xiong2022evolution, xiong2022collisional, liu2023systematic, liu2023universality, akaho2023collisional, shalgar2023neutrinos, kato2023collisional, froustey2024neutrino, wang2025effect} (FFI, SFI, and CFI, respectively).

Until several years ago, the predominant approach to this problem was to solve the exact quantum kinetic equations (QKEs) in simplified astrophysical models \cite{duan2010collective}. Increasing awareness of the prevalence of flavor instabilities gradually chipped away at this paradigm \cite{chakraborty2016collective, tamborra2021new}. Instabilities spontaneously break the artificial symmetries of simplified astrophysical models and generate small-scale spatial inhomogeneities in neutrino flavor. The formation of small-scale features in phase space, a generic trait also of weakly collisional plasmas and self-gravitating systems, poses a grave challenge to numerical simulation \cite{johns2020fast} (but see Refs.~\cite{shalgar2023neutrino, shalgar2024length, liu2025resolution} for debate over spatial resolution requirements and Refs.~\cite{myers2022neutrino, grohs2023neutrino, grohs2023twomoment, froustey2023neutrino, kneller2024quantum, froustey2024quantum, grohs2025advection} for work on quantum angular-moment closures as a way to alleviate the demands of momentum resolution).

Direct numerical simulation of the QKEs has largely been supplanted by approaches based on the asymptotic states that result from flavor instabilities in periodic-box calculations \cite{xiong2024robust, nagakura2024bhatnagar, zaizen2023simple, zaizen2025spectral, padillagay2025flavor, froustey2025predicting, liu2025asymptotic}. In the quantum Bhatnagar--Gross--Krook (BGK) method, for example, fast flavor conversion (FFC) is approximated as continuous relaxation toward a spatially averaged local asymptotic state \cite{nagakura2024bhatnagar}. Methods based on asymptotic states face their own limitations, however \cite{urquilla2025testing, johns2024subgrid}. While they are computationally tractable, it is questionable whether they adequately capture all of the oscillation physics. For instance, the newly discovered phenomenon of \textit{FFC without FFI}---where FFI creates small-scale inhomogeneities that go on to alter the flavor evolution after FFI itself is no longer operative---falls outside their scope \cite{urquilla2025testing}.

Over the past few years, two of us have developed an alternative picture in which neutrinos are in slowly evolving states of \textit{local mixing equilibrium}. Local equilibrium evolves due to astrophysical driving (collisions, advection over macroscopic distances, time evolution of the astrophysical fluid) and a kind of frictional response that stems from small-scale flavor fluctuations. The transport theory based on this picture is called \textit{miscidynamics}. Originally we introduced this theory under the approximation that the friction from subgrid fluctuations---what we later termed \textit{flavor-wave viscosity}---can be ignored \cite{johns2023thermodynamics, johns2024subgrid} (see Ref.~\cite{kost2025once} for numerical tests). Subsequently we appended a coarse-grained prescription for treating flavor-wave viscosity: a supplemental theory that we call \textit{flavor-wave transport} \cite{johns2025local}. Flavor-wave transport has been developed in parallel by Fiorillo and Raffelt over a series of papers \cite{fiorillo2024fast, fiorillo2025collective, fiorillo2026flavomon, fiorillo2026quasi}. Their work advances essentially the same set of ideas for approximating the dynamics of the small-scale fluctuations, often using the term \textit{flavomon} to refer to flavor waves and the label \textit{quasilinear theory} to refer to flavor-wave transport.\footnote{To be precise, the word \textit{flavomon} was introduced in Ref.~\cite{fiorillo2025collective} for the quanta of flavor waves, meant to be comparable to \textit{plasmon} and \textit{magnon} (the quanta of plasma waves and spin waves, respectively), although the authors of that work routinely use the term to refer to the standard non-quantized flavor waves. In our view, \textit{quasilinear} denotes a potentially useful approximation within flavor-wave transport theory, not a defining feature.}

In both formulations of flavor-wave transport, the one in Ref.~\cite{johns2025local} and the other in Refs.~\cite{fiorillo2024fast, fiorillo2025collective, fiorillo2026flavomon, fiorillo2026quasi}, flavor waves are quasiparticles whose properties are determined by the local mean background, whose interactions may be approximated using kinetic theory, and whose macroscopic propagation may be described using the ray approximation.\footnote{The concept of local mixing equilibrium, the defining characteristic of miscidynamics, is notably absent from Refs.~\cite{fiorillo2024fast, fiorillo2025collective, fiorillo2026flavomon, fiorillo2026quasi}. This is because the authors assume that the spatially averaged Hamiltonian vectors $\langle \bm{H}_{\bm{q}} \rangle$ and polarizations $\langle \bm{P}_{\bm{q}} \rangle$ are all aligned with the weak-interaction axis $\bm{z}$. Under this assumption, local mixing equilibrium [Eq.~\eqref{eq:mixeq}] trivially obtains. However, some flavor-mixing effects like MSW conversion \cite{wolfenstein1978neutrino, mikheyev1985resonance} and spectral swaps \cite{duan2006simulation, duan2006coherent, raffelt2007self2, raffelt2007adiabaticity} crucially involve deviations from such trivial configurations. See Ref.~\cite{kost2025once} for numerical examples of how flavor instability leads into adiabatic tracking of nontrivial mixing equilibria, quantitatively matching the predictions of miscidynamics. Contrary to the title of Ref.~\cite{fiorillo2025collective}, collective flavor conversion is \textit{not} always due to interaction of neutrinos with flavor waves. In some important cases it is due to the system moving through different mixing equilibria as parameters vary, as in spectral swaps \cite{johns2023thermodynamics}, relaxation in the once-in-a-lifetime encounter model \cite{kost2024once, kost2025once}, and at least some instances of CFI \cite{johns2023collisional}. In other cases collective flavor conversion reflects grid-level ($\bm{k} = \bm{0}$) collisionless instability, as in the motion of the slow \cite{hannestad2006self, duan2007analysis, johns2018strange} and fast \cite{johns2020neutrino, padilla2022neutrino, fiorillo2023slow} flavor pendula.}  This line of research builds on earlier work on flavor-wave properties that showed how to use linearized dispersion relations on static backgrounds to determine the energies $\Omega^{\textrm{R}} \equiv \textrm{Re}(\Omega)$ and growth/decay rates $\Omega^{\textrm{I}} \equiv \textrm{Im}(\Omega)$ of flavor waves, $\Omega$ being an eigenvalue of the linearized analysis \cite{banerjee2011linearized, izaguirre2017fast}. Flavor-wave transport leverages these properties, as well as some newer ones, to approximate the evolution of small-scale fluctuations. Among the more recent properties are the group velocities $\bm{v}^{\textrm{g}} \equiv \partial_{\bm{k}} \Omega^{\textrm{R}}$ and forces $\bm{F} \equiv -\partial_{\bm{r}} \Omega^{\textrm{R}}$ that accompany the reinterpretation of wave vector $\bm{k}$ as quasiparticle momentum \cite{johns2025local, fiorillo2026flavomon}. The additional properties of flavor-wave helicity and phase alignment \cite{liu2025dynamical}, which unlike the other properties mentioned concern the eigenstates rather than the eigenvalues, might be used to layer further approximations on flavor-wave transport, though as of yet they have not been utilized.

At the heart of flavor-wave transport is the expectation that flavor-wave eigensystems change slowly, on the scale of astrophysical variations, even if the flavor-wave states change quickly, on the scale of neutrino oscillations.\footnote{That flavor-wave eigensystems vary slowly is a postulate, and not an entirely obvious one. This expectation is backed up by the self-consistent picture advanced by miscidynamics. For eigensystems to vary on the astrophysical scale, the dynamics must exhibit two key features. First, the background polarizations $\langle \bm{P}_{\bm{q}} \rangle$ must be near local mixing equilibrium, since otherwise the precession terms $\langle \bm{H}_{\bm{q}} \rangle \times \langle \bm{P}_{\bm{q}} \rangle$ in the $\langle \bm{P}_{\bm{q}} \rangle$ equations of motion cause significant oscillation-scale evolution of the flavor-wave Hamiltonians $\mathcal{H}_{\bm{n}}$ (which depend on $\langle \bm{P}_{\bm{q}} \rangle$). Second, the flavor-wave viscosity---the backreaction of flavor waves on the background polarizations---must be small. This is ensured by the hypothesis that the viscosity is small whenever the system evolves close to stability (see Ref.~\cite{liu2025dynamical} for evidence that the viscosity vanishes at stability due to a combination of phase-aligned synchronization and effective phase randomization). A full articulation of this picture may be found in Refs.~\cite{johns2023thermodynamics, johns2024subgrid} and especially Ref.~\cite{johns2025local}, though a thorough understanding of these points is not required to appreciate the findings we report here.} The slow variation of eigensystems opens up potential for an adiabatic principle to apply to flavor-wave evolution, with the (probably important) eigenstate populations undergoing slow evolution even though the (hopefully inessential) phases do not. In the extreme limit, this principle is embodied by the \textit{quasistatic (QS) approximation} that the change in the background is so slow that it drives no transfer of population between energy levels.\footnote{We define adiabaticity for flavor waves to mean that the eigenstate populations are constant, hence so too is the flavor-wave entropy. Quasistaticity is a less stringent condition, as it pertains only to one mechanism of nonadiabaticity: transitions due to finite rates of change of the background. If we were to impose the adiabatic approximation on flavor waves, we would neglect not just nonquasistatic transitions, but also instability and damping ($\Omega^{\textrm{I}} \neq 0$) and nonlinear wave--wave interactions.}

In this paper, we evaluate the prospects of flavor-wave transport using numerical examples, with a particular focus on the critical notion of quasistaticity and the dispensable but convenient approximation of quasilinearity. 

As our study was nearing completion, Ref.~\cite{fiorillo2026quasi} appeared, presenting the very first numerical implementation of flavor-wave transport. In that work, Fiorillo and Raffelt demonstrate good agreement between the approximate and exact QKE solutions in sudden-instability test cases, where the initial neutrino distributions exhibit significant angular crossings and homogeneity apart from small seed fluctuations. Although this type of initial condition has been much studied, the relevance to real astrophysical environments is somewhat suspect \cite{johns2024subgrid, fiorillo2024fast}. The crucial feature lacking in sudden-instability test models is ongoing astrophysical driving. Based on Ref.~\cite{fiorillo2026quasi}, the success of flavor-wave transport is compelling for systems initialized in unstable, nearly homogeneous configurations that are allowed to equilibrate in isolation.

The natural next question is how flavor-wave transport fares in models where the conditions for instability are organically generated. Here we take this next step, exposing some of the important challenges that arise under slow driving and near-stable-equilibrium evolution. We will see in particular that another new property of flavor-wave eigenmodes, their penchant for coalescing at exceptional points, may in fact work against flavor-wave transport theory. We believe that this property, like phase alignment and helicity, has probably gone largely unnoticed because the far greater share of the literature has focused on flavor-wave eigenvalues, not the structure of the modes themselves. It has been known for some time that eigenvalues come in conjugate pairs in collisionless systems (\textit{e.g.}, see Ref.~\cite{morinaga2022fast}). This fact implies that the eigenvalues become degenerate as the system approaches marginal stability ($\Omega^{\textrm{I}} \rightarrow 0$). The fact that the eigenmodes become degenerate at the same time is a newer revelation \cite{johns2025local, liu2025dynamical} whose consequences for transport are unexplored.

We begin in Sec.~\ref{sec:theory} with a recap of flavor-wave transport theory using the formalism of Ref.~\cite{johns2025local}. We state more precisely the approximations that will be scrutinized in subsequent sections and develop the concept of \textit{flavor-wave parallel transport}. 

The focus of Sec.~\ref{sec:NL} is nonlinear wave--wave interactions, the terms that are neglected under the \textit{quasilinear (QL) approximation}. We show that the QL approximation has mixed success in periodic-box calculations with slow emission of neutrinos. The effect of the nonlinear coupling is wave turbulence, which redistributes energy and entropy over flavor waves at different Fourier modes $\bm{k}$ and energy bands $i$ \cite{johns2025neutrino, johns2025local, fiorillo2026quasi} (also see Ref.~\cite{mirizzi2015} for an earlier analogy with fluid turbulence). Errors accumulate in the QL evolution because the flavor-wave spectrum determines the flavor-wave viscosity, hence the response of the mean (spatially averaged) polarizations $\langle \bm{P}_{\bm{q}} \rangle$ to driving. If the nonlinear terms turn out to be important in astrophysical environments, this is not necessarily a dire conclusion because an approximate treatment already exists. In flavor-wave kinetics, wave--wave interactions are approximated as temporally coarse-grained collisionlike processes, mirroring the principles by which particle collisions emerge in the QKE from interactions among quantum fields \cite{johns2025local}. We leave validation of this proposal to future work.

In Sec.~\ref{sec:spec} we present numerical tests of the QS approximation. We show that it successfully applies to the flavor waves at some but not all wave vectors. As anticipated \cite{johns2025local, liu2025dynamical}, we find that exceptional points (\textit{i.e.}, spectral degeneracies where eigenstates coalesce) are endemic to evolution through marginally stable equilibria. Degeneracies at a particular wave vector $\bm{k}$ invalidate the QS approximation at that $\bm{k}$ by amplifying transitions between levels. This is similar to how a level crossing in Hermitian quantum mechanics undermines the adiabatic approximation. While in Hermitian quantum systems the Landau--Zener formula approximates the transition probability through an avoided level crossing, we are not aware of a comparable  analytic formula for level transitions through an exceptional point. Flavor-wave transport theory will need to overcome or somehow bypass this potentially severe obstacle. 

We turn in Sec.~\ref{sec:rot} to a question regarding the background polarizations and their influence on the flavor-wave spectra. In all of our numerical calculations, the mean polarizations $\langle \bm{P}_{\bm{q}} \rangle$ at all momenta $\bm{q}$ are oriented nearly along the weak-interaction axis $\bm{z}$. What happens to the flavor-wave spectra if rotational symmetry around $\bm{z}$ is spontaneously broken by the development of $\langle \bm{P}_{\bm{q}} \rangle^T \neq 0$ for some momenta, where $T$ denotes the part of the vector transverse to $\bm{z}$? It has become possible to address this question due to the recent generalization of linear stability analysis to encompass arbitrary background configurations \cite{johns2025local}. We present an example in which flavor waves that are stable on a rotationally symmetric background become unstable for small transverse values $\langle \bm{P}_{\bm{q}} \rangle^T \neq 0$. Importantly, we do not find that the fluctuating transverse parts make a difference to the overall evolution in our numerical calculations. Nonetheless, we cannot rule out the possibility that instabilities associated with \textit{broken rotational symmetry} influence the dynamics in other setups.

In Sec.~\ref{sec:disc} we summarize our findings and discuss approximations that are not tested in this work but may be crucial for the employment of flavor-wave transport in astrophysical simulations.

\section{Flavor-wave transport theory\label{sec:theory}} 

\subsection{Formalism}

Flavor waves are collective excitations of the neutrino medium analogous to spin waves, plasma waves, sound waves, and the like. Their ``on-shell'' properties are determined from the linearized eigensystem \cite{banerjee2011linearized, izaguirre2017fast}.\footnote{In the framework of Ref.~\cite{izaguirre2017fast}, the flavor-wave dispersion relation is obtained and solved under the assumptions that flavor coherence is small and the background polarizations are static and exclusively along $\bm{z}$. The first condition can be relaxed to the less stringent assumption of weak inhomogeneity, which merely requires that any individual $\bm{P}_{\bm{q},\bm{n} \neq \numz}$ is small, and the second can be loosened to allow for background polarizations in any state of mixing equilibrium \cite{johns2025local}. Arbitrary nonequilibrium backgrounds can also be adopted for linear analysis, but in this case the staticity of the background on short time scales is not ensured.} The eigenvalues reveal the energies $\Omega^{\textrm{R}}$ and growth/decay rates $\Omega^{\textrm{I}}$ as functions of the wave vector (flavor-wave momentum) $\bm{k}$. The eigenstates reveal additional properties like phase alignment and helicity \cite{liu2025dynamical}. The goal of flavor-wave transport is to leverage the local properties of flavor waves to approximate neutrino quantum kinetics. This is distinct from the more traditional use of these properties to identify instabilities in simulation data and toy models.

Flavor-wave transport hinges on flavor-wave properties varying slowly in space and time. This condition is made feasible by the paradigm of miscidynamics \cite{johns2023thermodynamics, johns2024subgrid}. Local mean polarizations evolve due to astrophysical driving and viscosity arising from subgrid fluctuations. The coarse-grained flavor evolution is anchored to the astrophysical scale.

Here we summarize and build on the formalism of Ref.~\cite{johns2025local}, where we developed flavor-wave transport as a way to approximate the evolution of subgrid fluctuations and thus approximate the flavor-wave viscosity. We start with the collisionless QKEs in the fast limit and with discrete momenta:
\begin{equation}
    \left(\partial_t + \vel \cdot \nabla \right) \pol_{\mo} = \mu \sum_{\mo'} (1 - \vel' \cdot \vel) \pol_{\mo'} \times \pol_{\mo},
    \label{eq:QKE}
\end{equation}
where $\pol_{\mo} = \pol_{\mo}(t, \pos)$ and $\mu \equiv \sqrt 2 G_F n_\nu$ is constant. We omit antineutrinos in this presentation of flavor-wave transport and in our later numerical tests, though they are straightforwardly incorporated. In a periodic cube with volume $\mathcal{V}$, the polarizations can be described by spatial Fourier modes
\begin{equation}
    \pol_{\mo, \num}(t) \equiv \frac{1}{\mathcal{V}} \int \dd^3 \pos \pol_{\mo}(t, \pos) e^{-i \wav \cdot \pos},
\end{equation}
which obey the Fourier-space QKE
\begin{equation}
    \label{eq:FourierEOM}
    \left(\partial_t + i \vel \cdot \wav \right) \pol_{\mo, \num} = \mu \sum_{\mo'} (1 - \vel' \cdot \vel) \sum_{\numm + \nummm = \num} \pol_{\mo', \numm} \times \pol_{\mo, \nummm}.
\end{equation}
In this work we solve for the evolution of the mean flavor polarizations $\langle\bm{P}_{\mo}\rangle = \pol_{\mo, \numz}$ by exactly solving Eq.~\eqref{eq:FourierEOM} with $\num = \numz$. While not crucial for our purposes here, this evolution can be recast in the form $\partial_t \langle \bm{P}_{\bm{q}} \rangle = - \gamma_{\bm{q}} \langle \bm{P}_{\bm{q}} \rangle$, where $\gamma_{\bm{q}}$ is the (not necessarily positive) flavor-wave viscosity arising from interactions between the mean polarizations and the subgrid inhomogeneities \cite{johns2025local}. Meanwhile, the Fourier modes at $\num \neq \bm{0}$ can be recast as wave functions
\begin{equation}
    \wf_{\num} \equiv \left(P^x_{\mo_1, \num}, P^y_{\mo_1, \num}, P^z_{\mo_1, \num}, \ldots, P^x_{\mo_N, \num}, P^y_{\mo_N, \num}, P^z_{\mo_N, \num} \right)^T
\end{equation}
with $3N$ components for $N$ discrete momenta. Each wave function obeys the Schr\"{o}dinger-like equation
\begin{equation}
    \label{eq:wfEOM}
    i \partial_t \wf_{\num} = \ham_{\num} \wf_{\num} + \nl_{\num},
\end{equation}
with Hamiltonian matrix $\ham_{\num}$ and $\wf_{\numm}$-dependent nonlinear term $\nl_{\num}$. In general, $\ham_{\num}$ is non-Hermitian. Explicitly, the Hamiltonian is
\begin{widetext}
\begin{equation}
    \ham_{\num} \equiv \begin{pmatrix}
        \hat \mo_1 \cdot \wav &
        -i H_{\mo_1, \numz}^z &
        i H_{\mo_1, \numz}^y &
        &
        0 &
        i \mu f_{1,N} \polscalar_{\mo_1, \numz}^z &
        -i \mu f_{1,N} \polscalar_{\mo_1, \numz}^y
        \\
        i H_{\mo_1, \numz}^z &
        \hat \mo_1 \cdot \wav &
        -i H_{\mo_1, \numz}^x &
        \cdots &
        -i \mu f_{1,N} \polscalar_{\mo_1, \numz}^z &
        0 &
        i \mu f_{1,N} \polscalar_{\mo_1, \numz}^x
        \\
        -i H_{\mo_1, \numz}^y &
        i H_{\mo_1, \numz}^x &
        \hat \mo_1 \cdot \wav &
        &
        i \mu f_{1,N} \polscalar_{\mo_1, \numz}^y &
        -i \mu f_{1,N} \polscalar_{\mo_1, \numz}^x &
        0
        \\
        &
        \vdots &
        &
        \ddots &
        &
        \vdots &

        \\
        0 &
        i \mu f_{N,1} \polscalar_{\mo_N, \numz}^z &
        -i \mu f_{1,N} \polscalar_{\mo_N, \numz}^y &
        &
        \hat \mo_N \cdot \wav &
        -i H_{\mo_N, \numz}^z &
        i H_{\mo_N, \numz}^y
        \\
        -i \mu f_{N,1} \polscalar_{\mo_N, \numz}^z &
        0 &
        i \mu f_{N,1} \polscalar_{\mo_N, \numz}^x &
        \cdots &
        i H_{\mo_N, \numz}^z &
        \hat \mo_N \cdot \wav &
        -i H_{\mo_N, \numz}^x
        \\
        i \mu f_{N,1} \polscalar_{\mo_N, \numz}^y &
        -i \mu f_{N,1} \polscalar_{\mo_N, \numz}^x &
        0 &
        &
        -i H_{\mo_N, \numz}^y &
        i H_{\mo_N, \numz}^x &
        \hat \mo_N \cdot \wav
    \end{pmatrix}
    ,
\end{equation}
\end{widetext}
where
\begin{equation}
    f_{i,j} \equiv 1 - \hat \mo_{i} \cdot \hat \mo_{j}
\end{equation}
and
\begin{equation}
    \bm{H}_{\mo_i, \numz} \equiv \mu \sum_{j} f_{j, i} \pol_{\mo_j, \numz},
\end{equation}
and the nonlinear term is defined by
\begin{equation}
    \nl_{\num, \ind} \equiv \sum_{\numm \notin \{\numz, \num\}} \wf_{\num - \numm}^T i V_{\ind} \wf_{\numm},
\end{equation}
where we also define the coupling matrices $V$, which, for integers $i$ labeling the components of the vectors $\Psi_{\bm{n}}$ and $N_{\bm{n}}$, are given by
\begin{equation}
    V_{3(\ind - 1) + 1} \equiv \mu \begin{pmatrix}
        & 0 & 0 & 0 & \\
        \cdots & 0 & 0 & f_{1,\ind} & \cdots \\
        & 0 & -f_{1,\ind} & 0 & \\
        & & \vdots & & \\
        & 0 & 0 & 0 & \\
        \cdots & 0 & 0 & f_{N,\ind} & \cdots \\
        & 0 & -f_{N,\ind} & 0 &
    \end{pmatrix}
    ,
\end{equation}
\begin{equation}
    V_{3(\ind - 1) + 2} \equiv \mu \begin{pmatrix}
        & 0 & 0 & -f_{1,\ind} & \\
        \cdots & 0 & 0 & 0 & \cdots \\
        & f_{1,\ind} & 0 & 0 & \\
        & & \vdots & & \\
        & 0 & 0 & -f_{N,\ind} & \\
        \cdots & 0 & 0 & 0 & \cdots \\
        & f_{N,\ind} & 0 & 0 &
    \end{pmatrix}
    ,
\end{equation}
and
\begin{equation}
    V_{3(\ind - 1) + 3} \equiv \mu \begin{pmatrix}
        & 0 & f_{1,\ind} & 0 & \\
        \cdots & -f_{1,\ind} & 0 & 0 & \cdots \\
        & 0 & 0 & 0 & \\
        & & \vdots & & \\
        & 0 & f_{N,\ind} & 0 & \\
        \cdots & -f_{N,\ind} & 0 & 0 & \cdots \\
        & 0 & 0 & 0 &
    \end{pmatrix}
    .
\end{equation}
Note that $\wf_{3(\ind - 1) + 1}$, $\wf_{3(\ind - 1) + 2}$, and $\wf_{3(\ind - 1) + 3}$ are equal to the $x$, $y$, and $z$ components, respectively, of $\pol_{\mo_\ind, \num}$. These definitions of the wave functions, Hamiltonians, and nonlinear terms are the same as those in Ref.~\cite{johns2025local} but without collisions.

When $\ham_{\num}$ is diagonalizable, we can write
\begin{equation}
    \vecs_{\num}^{-1} \ham_{\num} \vecs_{\num} = \vals_{\num} \quad \text{($\vals_{\num}$ diagonal)},
    \label{eq:diagonalize}
\end{equation}
where the columns of $\vecs_{\num}$ are eigenvectors of $\ham_{\num}$ and $\vals_{\num}$ contains their eigenvalues. Transforming into the eigenbasis, we define
\begin{equation}
    \co_{\num} = \vecs_{\num}^{-1} \wf_{\num},
\end{equation}
which evolves according to
\begin{equation}
    i \partial_t \co_{\num} = \vals_{\num} \co_{\num} + \vecs_{\num}^{-1} \nl_{\num} - i S_{\num}^{-1} (\partial_t S_{\num}) \co_{\num}.
    \label{eq:PhiEOM}
\end{equation}
The terms on the righthand side of Eq.~\eqref{eq:PhiEOM} correspond to three physically distinct processes:
\begin{align}
    \vals_{\num} \co_{\num}: &~\textrm{Flavor-wave phase evolution ($\Omega^{\textrm{R}}$)} \notag\\
    &~\textrm{and growth/decay ($\Omega^{\textrm{I}}$) due to} \notag\\
    &~\textrm{interaction with the background}\notag\\
    &~\textrm{polarizations } \langle\bm{P}_{\bm{q}}\rangle. \notag\\
    \vecs_{\num}^{-1} \nl_{\num} : &~\textrm{Nonlinear interactions among flavor}\notag\\
    &~\textrm{waves.}\notag\\
    i S_{\num}^{-1} (\partial_t S_{\num}) \co_{\num} : &~\textrm{Transitions of flavor waves at wave}\notag\\
    &~\textrm{number $\bm{n}$ between the energy levels}\notag\\
    &~\textrm{$\Omega^{\textrm{R}}_{\bm{n},i}$ driven by a finite rate of}\notag\\
    &~\textrm{change of the Hamiltonian $\mathcal{H}_{\bm{n}}$.}\notag
\end{align}
The approximations we test in this paper pertain to the last two terms.

\subsection{The quasilinear approximation}

The QL approximation neglects the second-to-last term, imposing
\begin{equation}
    N_{\bm{n}} = 0 ~~~~ \textrm{(quasilinear approximation)}\label{eq:QLapprox}
\end{equation}
in Eq.~\eqref{eq:PhiEOM}. In the kinetic theory of flavor-wave interactions, $N_{\bm{n}}$ consists of scattering events among flavor waves that redistribute flavor-wave energy $\Omega^{\textrm{R}}$ and momentum $\bm{k}$ \cite{johns2025local}. As we have said, we are here testing whether $N_{\bm{n}}$ can safely be ignored in its entirety. We are not testing the kinetic approximation.

\subsection{Parallel transport}

To address the final term in Eq.~\eqref{eq:PhiEOM}, we need to introduce the notion of flavor-wave parallel transport. Let $\rvec_{\num, \ind}$ denote the $\ind$th column of $\vecs_{\num}$, \textit{i.e.}, the $\ind$th right eigenvector of $\ham_{\num}$, and let $\lvec_{\num, \ind}^\conj$ denote the $\ind$th row of $\vecs_{\num}^{-1}$, \textit{i.e.}, the Hermitian conjugate of the $\ind$th left eigenvector of $\ham_{\num}$. Note that
\begin{equation}
    \vecs_{\num}^{-1} \vecs_{\num} = 1 \implies \lvec_{\num, \ind}^\conj \rvec_{\num, \indd} = \delta_{\ind \indd}.
\end{equation}
We can now write the matrix factor in the last term of Eq.~\eqref{eq:PhiEOM} as
\begin{equation}
    [S_{\num}^{-1} (\partial_t S_{\num})]_{\ind \indd} = \lvec_{\num, \ind}^\conj (\partial_t \rvec_{\num, \indd}).
    \label{eq:transport-eig}
\end{equation}
For simplicity, assume for the moment that the eigenvalues of $\ham_{\num}$ are nondegenerate. Then the (right) eigenvectors defined by Eq.~\eqref{eq:diagonalize} are each determined only up to an arbitrary magnitude and phase. To be explicit, we can choose different magnitudes (scaled by $\magni_{\ind}$) and phases (shifted by $\phase_{\ind}$) via the transformation
\begin{equation}
    \rvec_{\num, \ind} \to \magni_{\ind} e^{i \phase_{\ind}} \rvec_{\num, \ind},
\end{equation}
where $\magni_{\ind}$ and $\phase_{\ind}$ are real numbers that may depend on time. For brevity, we omit the subscript $\num$ on them. Under this transformation, we have
\begin{equation}
     \lvec_{\num, \ind}^\conj \to \left(\magni_{\ind} e^{i \phase_{\ind}} \right)^{-1} \lvec_{\num, \ind}^\conj
\end{equation}
and
\begin{align}
    [S_{\num}^{-1} (\partial_t S_{\num})]_{\ind \ind} \to & [S_{\num}^{-1} (\partial_t S_{\num})]_{\ind \ind} + \frac{\partial_t \magni_{\ind}}{\magni_{\ind}} + i \partial_t \phase_{\ind}
    \label{eq:transformed}
\end{align}
for each $i$. Equation~\eqref{eq:transformed} shows that the freedom to choose the magnitudes and phases of the eigenvectors (or more precisely their time derivatives) via $\magni_{\ind}$ and $\phase_{\ind}$, respectively, is equivalent to the freedom to choose the real and imaginary parts, respectively, of the diagonal elements of $\vecs_{\num}^{-1} (\partial_t \vecs_{\num})$. Setting all of the diagonal elements of $\vecs_{\num}^{-1} (\partial_t \vecs_{\num})$ to zero defines parallel transport:
\begin{equation}
    [\vecs_{\num}^{-1} (\partial_t \vecs_{\num})]_{\ind \ind} = 0 ~~~ \textrm{(parallel transport).}
\end{equation}
No sum over $\ind$ is implied.

This concept may be more familiar when the Hamiltonian is Hermitian. In this case, $\lvec_{\num, \ind} \propto \rvec_{\num, \ind}$ with a real proportionality constant, and using Eq.~\eqref{eq:transport-eig}, we have
\begin{equation}
    \text{Re}\left([S_{\num}^{-1} (\partial_t S_{\num})]_{\ind \ind} \right)
    \propto \text{Re}\left(\rvec_{\num, \ind}^* \partial_t \rvec_{\num, \ind} \right)
    \propto \partial_t \abs{\rvec_{\num, \ind}}^2.
\end{equation}
So, in the Hermitian case, setting the real parts of the diagonal elements of $S_{\num}^{-1} (\partial_t S_{\num})$ to zero enforces that the magnitudes of the eigenvectors remain constant in time. Meanwhile, the imaginary parts produce geometric phases, and setting them to zero enforces parallel transport of the phases of the eigenvectors. In the non-Hermitian case, the magnitudes of the eigenvectors are no longer constant under parallel transport, and instead, like the phases, their evolution generally depends on the path taken by the Hamiltonian in its parameter space. Thus, for flavor waves as for other non-Hermitian quantum systems, there arise complex geometric phases \cite{dattoli1990geometrical, mehri2008geometric}, which differ from the types of geometric phases previously considered for neutrinos \cite{he2005berry, johns2017geometric, johns2022geometric}.

\subsection{The quasistatic approximation}

While parallel transport dictates that the diagonal elements of $S_{\num}^{-1} (\partial_t S_{\num})$ vanish, the off-diagonal elements cannot be arbitrarily chosen because they drive physical transitions among flavor-wave energy levels. However, in the QS limit, in which background parameters (\textit{i.e.}, all those appearing in $\mathcal{H}_{\bm{n}}$) change infinitely slowly, the off-diagonal elements vanish as well. 

The QS limit can be understood as follows. Let $T$ be the long astrophysical timescale over which the Hamiltonian changes by a relative amount of order unity: $(\partial_t \ham_{\num}) / \ham_{\num} \sim 1 / T$. When terms in the equation of motion that are of order $1 / T^2$ or smaller are integrated over the timescale $T$, they lead to changes in the state of order $1 / T$ or smaller, which vanish as $T \to \infty$. Suppose the Hamiltonian is Hermitian, and that its eigenvalues are significantly nondegenerate:
\begin{equation}
    \abs{\vals_{\num, \ind} - \vals_{\num, \indd}} \gg \frac{1}{T} \quad (i \neq j).
    \label{eq:Hermitian-condition}
\end{equation}
Then the first term on the righthand side of Eq.~\eqref{eq:PhiEOM}, unlike the third, is not suppressed by $1/T$. Assuming the second term $\nl_{\num}$ is also small, the unsuppressed evolution of $\co_{\num}$ is that its $\ind$th component oscillates at the rate $\vals_{\num, \ind}$. Meanwhile, each off-diagonal matrix element $[\vecs_{\num}^{-1} (\partial_t \vecs_{\num})]_{\ind \indd}$ in the third term couples different components $\ind$ and $\indd$ of $\co_{\num}$. Their phase difference oscillates rapidly compared to any other evolution of $\co_{\num}$ [Eq. \eqref{eq:Hermitian-condition}], leading to an additional suppression of these couplings by a factor $(\abs{\vals_{\num, \ind} - \vals_{\num, \indd}} T)^{-1}$. In other words, in Eq. \eqref{eq:PhiEOM}, we effectively have
\begin{equation}
    [\vecs_{\num}^{-1} (\partial_t \vecs_{\num})]_{\ind \indd} \propto \frac{1}{T^2} \quad (i \neq j),
\end{equation}
and, even after Eq. \eqref{eq:PhiEOM} is integrated over $T$, the contributions from these off-diagonal elements to the evolution of $\co_{\num}$ vanish as $T \to \infty$. The requirement in this derivation that the eigenvalues are significantly degenerate may be generalized to the case of non-Hermitian $\ham_{\num}$ \cite{ibanez2014adiabaticity}.

Under parallel transport (which we henceforth assume) and the QS approximation, the last term in Eq.~\eqref{eq:PhiEOM} drops out completely:
\begin{equation}
    i S_{\num}^{-1} (\partial_t S_{\num}) \co_{\num} = 0 ~~~ \textrm{(quasistatic approximation)}. \label{eq:QSapprox}
\end{equation}
Essentially the same argument as above explains why quantum systems can evolve adiabatically over arbitrarily long times. However, we deliberately draw a distinction in our terminology between quasistaticity and adiabaticity. We use the latter term to describe flavor-wave evolution with no changes in the populations $|\co_{\num,\ind}|$, hence no entropy transfer among the flavor waves or between the flavor waves and the mean polarizations. Nonadiabaticity may stem from nonquasistaticity (finite rates of change), but it may also stem from instabilities ($\Omega^{\textrm{I}} \neq 0$, in which case the entropy transfer involves the mean polarizations) or wave--wave interactions ($\nl_{\num} \neq 0$). In standard Hermitian quantum systems, the concepts of nonquasistaticity and nonadiabaticity are equivalent, hence the temptation to think of Eq.~\eqref{eq:QSapprox} as an adiabatic approximation. We find it useful to resist this temptation and disentangle the two.

\section{Nonlinear wave--wave coupling\label{sec:NL}}

Now we turn to the numerical tests. The exact QKE simulations we present here and in later sections were carried out using a modified version of the open-source code NuGas developed by Duan.\footnote{\href{https://github.com/UNM-NuCO/nugas}{https://github.com/UNM-NuCO/nugas}.}

\begin{figure*}
    \centering
    \includegraphics[width=0.45\linewidth]{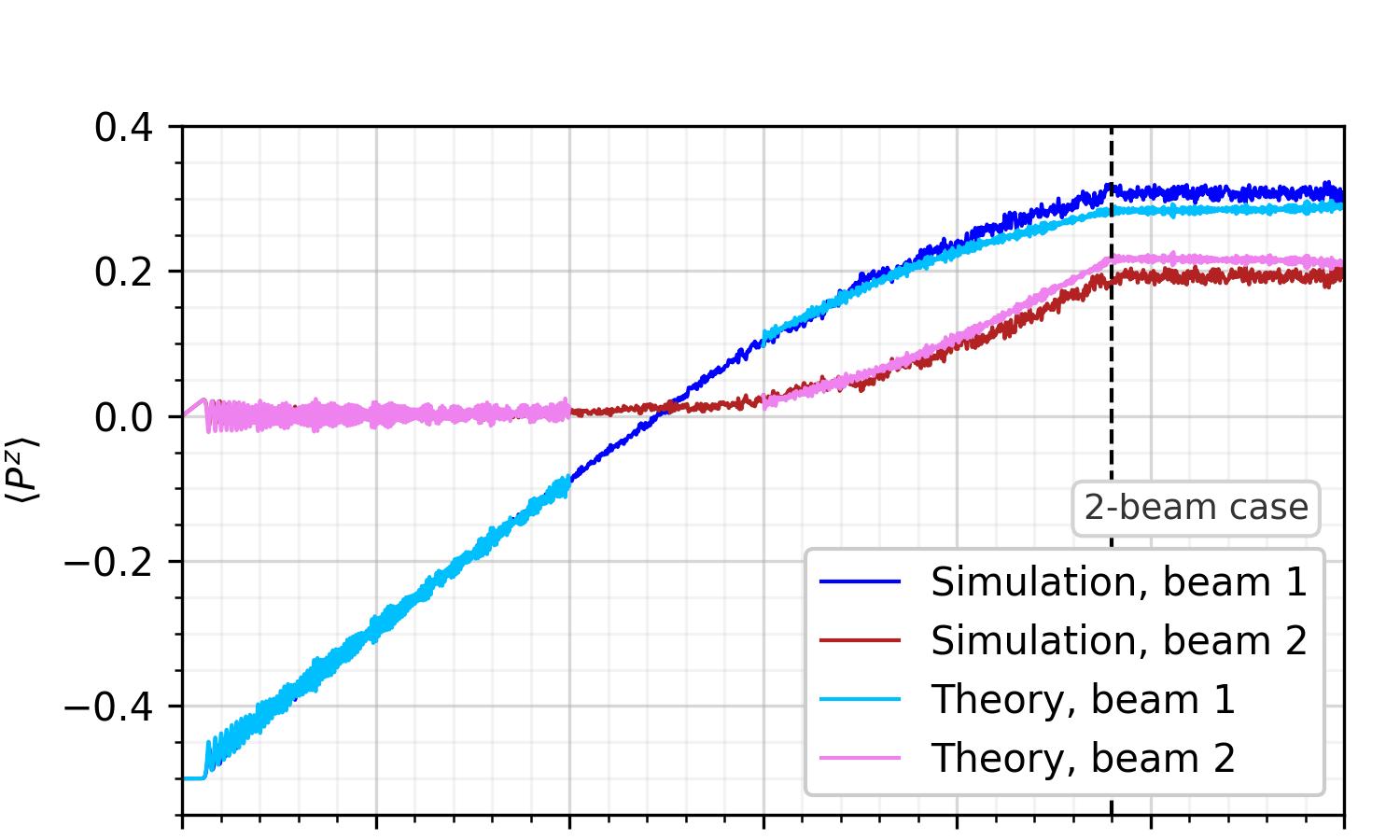}
    \includegraphics[width=0.45\linewidth]{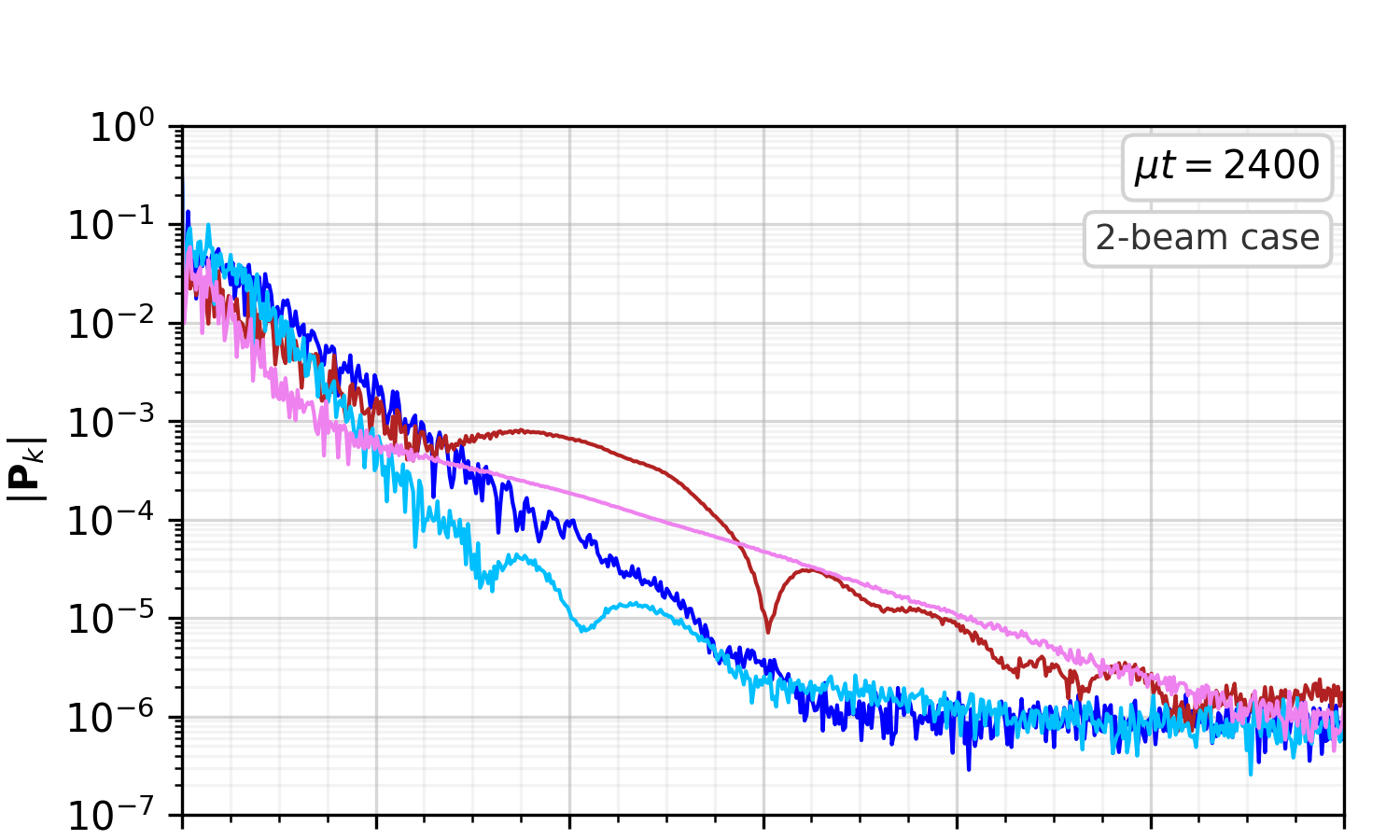}
    \\
    \includegraphics[width=0.45\linewidth]{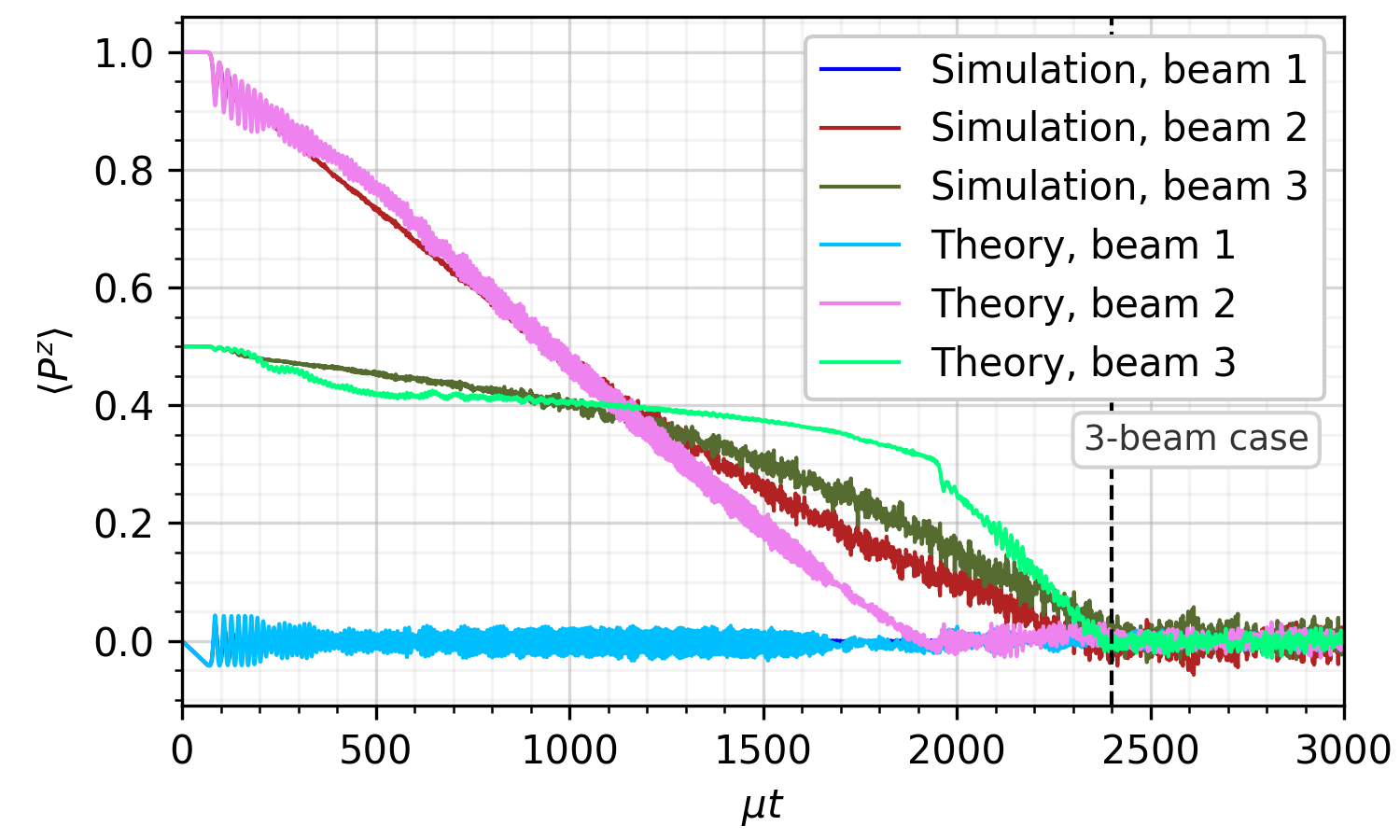}
    \includegraphics[width=0.45\linewidth]{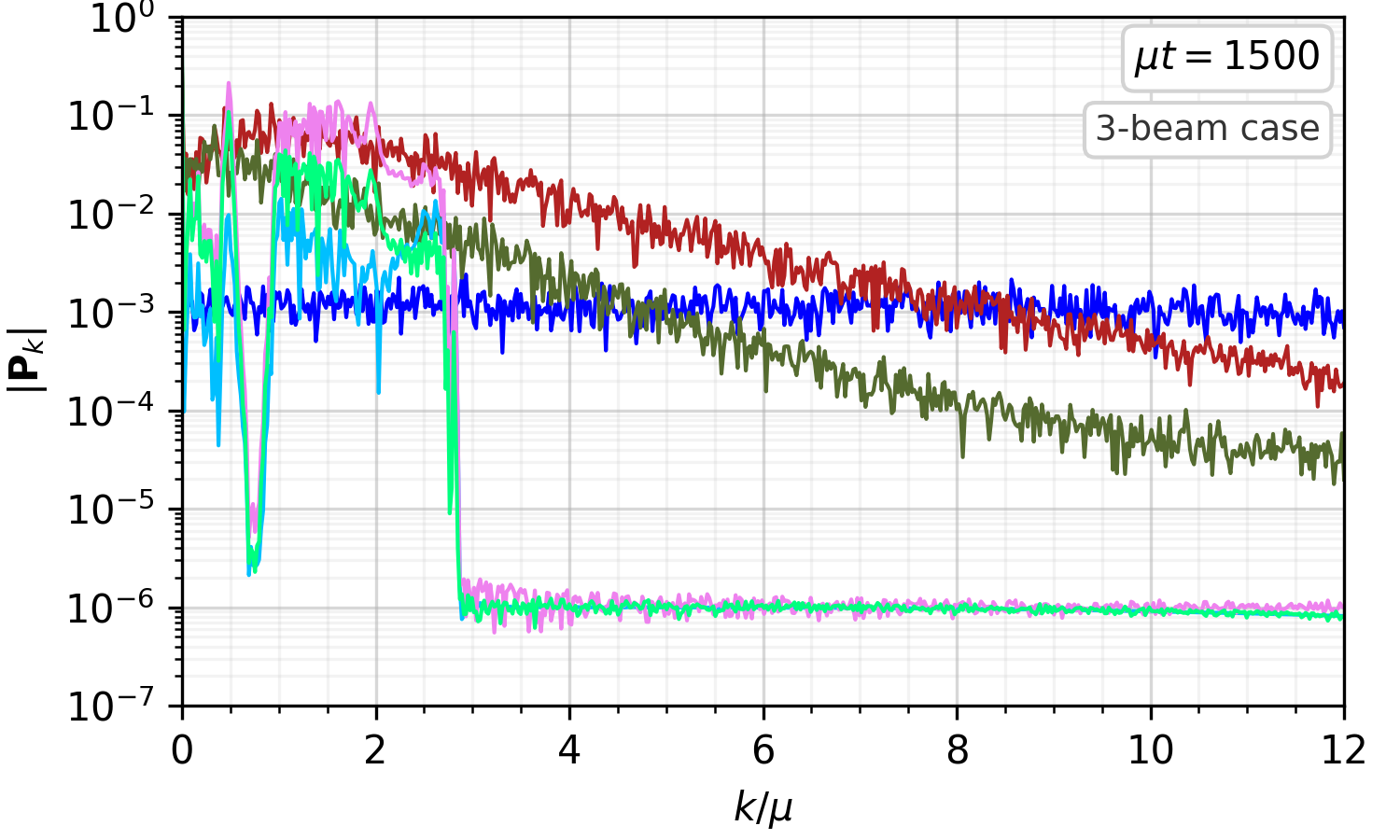}
    
    \caption{Comparison between simulation (exact QKE, solid curves) and theory (QL approximation, dashed curves) in the two-beam (upper panels) and three-beam (lower panels) test cases described in the main text, showing mean polarizations $\langle P^z \rangle$ vs. time $\mu t$ (left) and Fourier-mode magnitudes $|\bm{P}_{k}|$ vs. wave vector $k/\mu$ at select times (right). The vertical dashed lines in the left panels indicate the time when neutrino injection stops ($\mu t = 2400$). In the two-beam case, we resume the QL evolution after the beam crossing using the exact solution at $\mu t = 1500$. This choice exaggerates the agreement because discrepancies tend to accumulate over time. While the mean polarizations in the QL solution are in good agreement with those in the exact solution throughout the two-beam calculation, in the three-beam scenario they deviate noticeably during the second half of injection. These deviations arise from the significant discrepancies in $\bm{P}_{k \neq 0}$ due to the omission of nonlinear wave--wave interactions.}
    \label{fig:QL}
\end{figure*}

In this section we examine the successes and limitations of the QL approximation, which neglects nonlinear wave--wave coupling [Eq.~\eqref{eq:QLapprox}]. The flavor-wave equation of motion under the QL approximation is
\begin{equation}
    i \partial_t \co_{\num} = \vals_{\num} \co_{\num} - i S_{\num}^{-1} (\partial_t S_{\num}) \co_{\num},
    \label{eq:PhiEOMQL}
\end{equation}
although in our numerical calculations we evolve $\bm{P}_{\bm{q},\bm{n}}$ directly rather than adopt the time-dependent eigenbasis. We show Eq.~\eqref{eq:PhiEOMQL} here only as a reminder of the remaining physical elements.

Our test cases are similar to those studied in Refs.~\cite{fiorillo2024fast, liu2024quasi, urquilla2025testing}. Neutrinos are slowly injected into one of two or three momentum beams. The injection is uniform over the periodic box and constant in time over $\mu t \in [0, 2400]$. In the two-beam case, $\nu_e$ is injected into beam 2. In the three-beam case, $\nu_\mu$ is injected into beam 1. Explicitly, the injection is defined by adding the term $\zeta \mu / 2400$ to the righthand side of the equation of motion for $\partial_t \polscalar_{\mo_\text{injected}, \numz}^z$ [Eq. \eqref{eq:FourierEOM}] while $0 \leq \mu t \leq 2400$, and 0 otherwise, where $\mo_\text{injected}$ labels the momentum of the beam being injected into, and $\zeta$ is a dimensionless constant. In the two-beam case, $\zeta = 1$, while in the three-beam case, $\zeta = -3 / 2$. The simulations are run to a final time of $\mu t = 3000$.

Figure~\ref{fig:QL} compares the results from the simulation (the exact QKE) and theory (the QL approximation applied to the QKE). The left panels show the spatial averages $\langle P^z \rangle$ for the individual beams as functions of time $\mu t$. The right panels show the subgrid quantities $| \bm{P}_{k} |$ for the individual beams at particular snapshot times ($\mu t = 2400$ in the two-beam case and $\mu t = 1500$ in the three-beam case) as functions of wave vector $k / \mu$. In the theory solutions, we begin at $t = 0$ using the same initial conditions as in the simulation. In the two-beam case, we halt the QL evolution at $\mu t = 1000$ and resume again at $\mu t = 1500$. When resuming the theory evolution, we restart using the $\mu t = 1500$ data from the simulation, so that the simulation and theory are forced to agree at this time. The choice to skip over $\mu t \in [1000, 1500]$ in the two-beam case is based on the fact that both beams have $\langle P^z \rangle \approx 0$ in this interval. The theory calculation in the three-beam case runs continuously over $\mu t \in [0, 3000]$.

The agreement in $\langle P^z \rangle$ between simulation and theory is decent in the two-beam case, particularly for $\mu t \in [0, 1000]$. It is nonetheless concerning that discrepancies continuously accumulate for $\mu t \in [1500, 3000]$. Moreover, at $\mu t = 1500$, the theory solution inherits the flavor wave information from the exact solution as explained above. This choice exaggerates the agreement between theory and simulation. The agreement in the three-beam case is less compelling than in the two-beam case. Although the qualitative behavior is largely correct, quantitative discrepancies are quite apparent. 

Discrepancies in $\langle P^z \rangle$ are reflective of differences in the $| \bm{P}_{k} |$ Fourier spectra, as shown in the right panels of Fig.~\ref{fig:QL}. In the two-beam case, the agreement between theory and simulation is reasonably good across all $k / \mu$ plotted. We stress once more that the two-beam QL solution inherits full $\bm{P}_{k}$ information from the simulation at $\mu t = 1500$. In the three-beam case, the disagreement is much more significant.

The $| \bm{P}_{k} |$ spectra in the QL solutions are formed by FFI. In the lower right panel of Fig.~\ref{fig:QL}, all of the theory curves drop off to $| \bm{P}_k | \approx 10^{-6}$ at $k / \mu \gtrsim 2.8$ because $10^{-6}$ is the initial seed value of the subgrid fluctuations and at no point in the evolution are there unstable modes with $k / \mu \gtrsim 2.8$. In the full QKE solution, nonlinear wave--wave interactions couple flavor waves at different wave vectors and thus redistribute $| \bm{P}_k |$ over $k$. This is flavor-wave turbulence \cite{johns2025neutrino, johns2025local, fiorillo2026quasi}. In many instances it leads to exponential decline of $|\bm{P}_k|$ as a function of $k$ (see, \textit{e.g.}, Ref.~\cite{richers2022code}). We have speculated elsewhere that this exponential scaling may reflect at least partial thermalization (entropy-maximization) of the flavor-wave spectrum \cite{johns2025local}.

The validity of the QL approximation is highly dependent on the model and the duration of evolution. Its applicability in realistic environments is unclear and cannot be satisfactorily assessed by periodic-box calculations. In reality, flavor waves propagate both into and out of any small region in a supernova or merger. They are constantly produced in marginally stable regions that are being driven toward instability (\textit{i.e.}, regions at the edge of instability) \cite{fiorillo2024fast}. They can be both emitted and absorbed in regions where there is FFC without FFI or variants of this phenomenon \cite{urquilla2025testing}. In particular, flavor waves are depleted from beam 1 in the two-beam case at $\mu t > 1200$, causing the repolarization of this beam---\textit{kinematic recoherence}, to be contrasted with the much more familiar kinematic decoherence \cite{raffelt2007self}. While we cannot conclusively say whether the QL approximation is acceptable for astrophysical simulations, it is not obviously reliable for the types of calculations presented here. This finding motivates the testing of flavor-wave kinetics, which removes rapidly varying phases in the nonlinear terms by dropping nonresonant wave--wave interactions \cite{johns2025local}.

\section{Spectral degeneracies\label{sec:spec}} 

Next we demonstrate a potentially significant challenge to flavor-wave transport that was anticipated in Refs.~\cite{johns2025local, liu2025dynamical}: exceptional points (\textit{i.e.}, eigenmode degeneracies) in the flavor-wave spectra entail significant nonquasistatic level transitions.

The $\co_{\num}$ equations of motion [Eq.~\eqref{eq:PhiEOM}] are equivalent to the $\pol_{\mo}$ equations of motion [Eq.~\eqref{eq:QKE}] as long as $\ham_{\num}$ is diagonalizable at all $\num$. If the Hamiltonian evolves smoothly but loses its diagonalizability at time $t$, a pair of its eigenvectors and their eigenvalues come together at an exceptional point. Since the decomposition of the state $\wf_{\num}$ into eigenvector components breaks down near exceptional points, $\co_{\num}$ cannot be evolved continuously across an exceptional point using Eq.~\eqref{eq:PhiEOM}. Moreover, in the vicinity of an exceptional point, the QS approximation breaks down due to near-degeneracies in the flavor-wave spectrum. 

Exceptional points are characteristic of marginally stable equilibria. Consider the behavior of conjugate modes on either side of the stability threshold: on the unstable side, one mode grows with rate $\Omega^{\textrm{I}}$ and the other decays at the same rate.\footnote{$\mathcal{PT}$ symmetry is often the reason non-Hermitian Hamiltonians have either all real eigenvalues or spectra with complex-conjugate pairs. Whether some abstract $\mathcal{PT}$ symmetry accounts for this behavior in collisionless neutrino flavor evolution has not yet been explored.} At marginal stability, these modes fuse together at an exceptional point. Fluctuations in a neutrino system near marginal stability thus cause continual merging and splitting of modes. Our numerical results bear out this expectation.

In our numerical calculations, in order to characterize when the eigensystem of a given $\num$ is sufficiently nondegenerate for the QS approximation to be applicable, we use the simple criterion
\begin{equation}
    \min_{i, j; i \neq j} \abs{\vals_{\num, i} - \vals_{\num, j}} > \thresh, \label{eq:QScriterion}
\end{equation}
where $C$ is some threshold that may be varied. Note that this criterion is defined independently for each $\num$, and is defined even when the eigenvalues $\vals_{\num, \ind}$ are complex rather than real. A more rigorous quasistatic criterion could be adapted from Ref.~\cite{ibanez2014adiabaticity}, which examined the conditions under which adiabaticity applies to the evolution of non-Hermitian quantum systems. Numerically implementing a more careful criterion is irrelevant for our purposes here. It would not change our overall conclusion that quasistaticity is subverted by spectral degeneracies.

In our calculations, we use a hybrid method that falls back on exactly solving the $\Psi_{\bm{n}}$ equation of motion [Eq.~\eqref{eq:wfEOM}] whenever the quasistatic criterion given by Eq.~\eqref{eq:QScriterion} is not met. In effect, we abandon favor-wave transport and return to solving the full QKE evolution of $\bm{P}_{\bm{q},\bm{n}}$ anytime quasistaticity is suspected to fail at that $\bm{n}$. This is a hybrid method because we continue to apply the QS approximation and solve 
\begin{equation}
    i \partial_t \co_{\num} = \vals_{\num} \co_{\num} + \vecs_{\num}^{-1} \nl_{\num}.
    \label{eq:PhiEOMQS}
\end{equation}
for all $\bm{n}$ that do satisfy Eq.~\eqref{eq:QScriterion}.

We show pseudocode in Algorithm~\ref{alg:hybrid} that captures the main aspects of the numerical procedure. For simplicity, this presents the Euler method, although in our code we use RK4. We use a function called \texttt{match} to ensure that the eigenvector coefficients at a new time are associated with the correct eigenvectors at the previous time. This is done based on the overlap between new and old eigenvectors. In our calculations, we set the $\bm{n} = \bm{0}$ vectors to point precisely along $\pm \bm{z}$, approximating as zero the small, fluctuating transverse parts. We find that this helps prevent numerical instability by ensuring that the eigenmode tracking is done correctly. Note that keeping track of the eigenvector evolution is essential because it allows us to convert between $\co_{\num}$ and $\wf_{\num}$ using $\vecs_{\num}$.

\begin{algorithm}
\caption{Hybrid method}\label{alg:hybrid}
\begin{algorithmic}

\State $t \gets 0$  \Comment{Set initial quantities.}
\State $\pol_{\mo, \numz}(0) \gets \dots$
\State $\wf_{\num}(0) \gets \dots$
\State $\co_{\num}(0) \gets \vecs_{\num}^{-1}(0) \wf_{\num}(0)$
\State $t_f \gets \dots$  \Comment {Set the final time.}
\State $\Delta t \gets \dots$  \Comment {Set the time step size.}
\State $\thresh \gets \dots$  \Comment {Set the quasistatic criterion threshold.}

\While{$t < t_f$}  \Comment {The integration loop.}
    \For{each $\nums$}
        \State $\pol_{\mo, \numz}(t + \Delta t) \gets -i \nl_{\mo, \numz}(t) \Delta t$  \Comment{Evolve $\pol_{\mo, \numz}$.}
        \If{$c(\num, \thresh)$}  \Comment{Evolve $\co_{\num}$ with the QS approx.}
            \State $\co_{\num}(t + \Delta t) \gets -i \vals_{\num}(t) \co_{\num}(t) \Delta t$
            \State $\wf_{\num}(t + \Delta t) \gets \vecs_{\num}(t + \Delta t) \co_{\num}(t + \Delta t)$
        \Else  \Comment{Evolve $\wf_{\num}$ without the QS approx.}
            \State $\wf_{\num}(t + \Delta t) \gets -i \left(\ham_{\num}(t) \wf_{\num}(t) + \nl_{\num}(t) \right) \Delta t$
            \State $\co_{\num}(t + \Delta t) \gets \vecs_{\num}^{-1}(t + \Delta t) \wf_{\num}(t + \Delta t)$
        \EndIf
    \EndFor
    \State $t \gets t + \Delta t$
\EndWhile

\vspace{5 pt} \hrule \vspace{5 pt}

\Function{$c$}{$\num, \thresh$}  \Comment{The quasistatic criterion.}
    \If{$\min_{\ind, \indd; \ind \neq \indd}\left(\abs{\vals_{\num, \ind} - \vals_{\num, \indd}} \right) > \thresh$}
        \State \Return \texttt{true}
    \Else
        \State \Return \texttt{false}
    \EndIf
\EndFunction

\vspace{5 pt} \hrule \vspace{5 pt}

\Function{$(\vecs_{\num}, \vals_{\num})$}{t}  \Comment{Calculate the eigensystem.}
    \State $\vecs_{\num}(t), \vals_{\num}(t) \gets \texttt{diagonalize}\left(\ham_{\num}\left(\pol_{\mo, \numz}(t) \right) \right)$
    \If{$t > 0$}
        \State $\vecs_{\num}(t), \vals_{\num}(t) \gets \texttt{match}\left(\vecs_{\num}(t), \vals_{\num}(t), \vecs_{\num}(t - \Delta t) \right)$
        \State $\vecs_{\num}(t) \gets \texttt{paralleltransport}\left(\vecs_{\num}(t), \vecs_{\num}(t - \Delta t) \right)$
    \EndIf
    \State \Return $\vecs_{\num}(t), \vals_{\num}(t)$
\EndFunction

\vspace{5 pt} \hrule \vspace{5 pt}

\Function{match}{$\vecs_{\num}(t), \vals_{\num}(t), \vecs_{\num}(t - \Delta t)$}
    \For{each $\ind$}
        \State $\indd \gets \max_{\indd}\left(\lvec_{\num, \ind}^\conj(t - \Delta t) \rvec_{\num, \indd}(t) \right)$
        \State $\rvec_{\num, \ind; \text{new}}(t), \vals_{\num, \ind; \text{new}}(t) \gets \rvec_{\num, \indd}(t), \vals_{\num, \indd}(t)$
    \EndFor
    \State \Return $\vecs_{\num; \text{new}}(t), \vals_{\num; \text{new}}(t)$
\EndFunction

\vspace{5 pt} \hrule \vspace{5 pt}

\Function{paralleltransport}{$\vecs_{\num}(t), \vecs_{\num}(t - \Delta t)$}
    \For{each $\ind$}
        \State $\rvec_{\num, \ind}(t) \gets \rvec_{\num, \ind}(t) / \left(\lvec_{\num, \ind}^\conj(t - \Delta t) \rvec_{\num, \ind}(t) \right)$
    \EndFor
    \State \Return $\vecs_{\num}(t)$
\EndFunction

\end{algorithmic}
\end{algorithm}

\begin{figure*}
    \centering
    \includegraphics[width=0.9\linewidth]{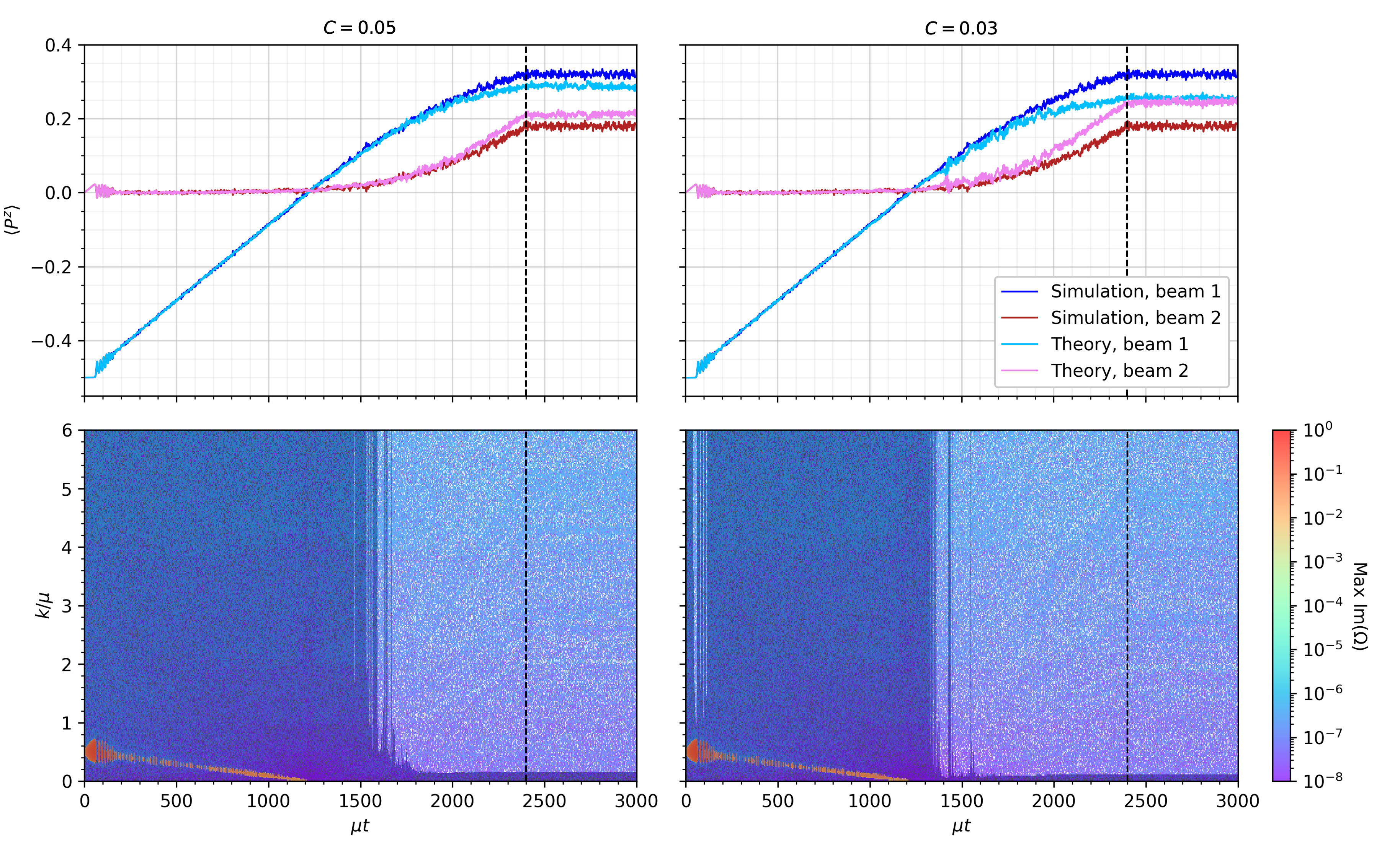}
    
    \caption{Comparison of simulation (exact QKE) and theory (hybrid method selectively applying the QS approximation) for the two-beam test case, showing the mean polarizations $\langle P^z \rangle$ (upper panels) for two different QS thresholds $C = 0.05$ (left) and $C = 0.03$ (right). The definition of $C$ is found in Eq.~\eqref{eq:QScriterion}. In the lower panels, shaded regions show the wave vectors $k / \mu$ that the QS approximation does not apply to at time $\mu t$. The colors indicate the maximum growth rates $\Omega^{\textrm{I}}$ obtained from the eigenanalysis in the hybrid calculation. The accuracy of the hybrid calculation breaks down for $C \lesssim 0.05$ as the QS approximation is erroneously applied to wave vectors at which flavor waves undergo significant nonadiabatic transitions.}
    \label{fig:QS2beam}
\end{figure*}

\begin{figure*}
    \centering
    \includegraphics[width=0.9\linewidth]{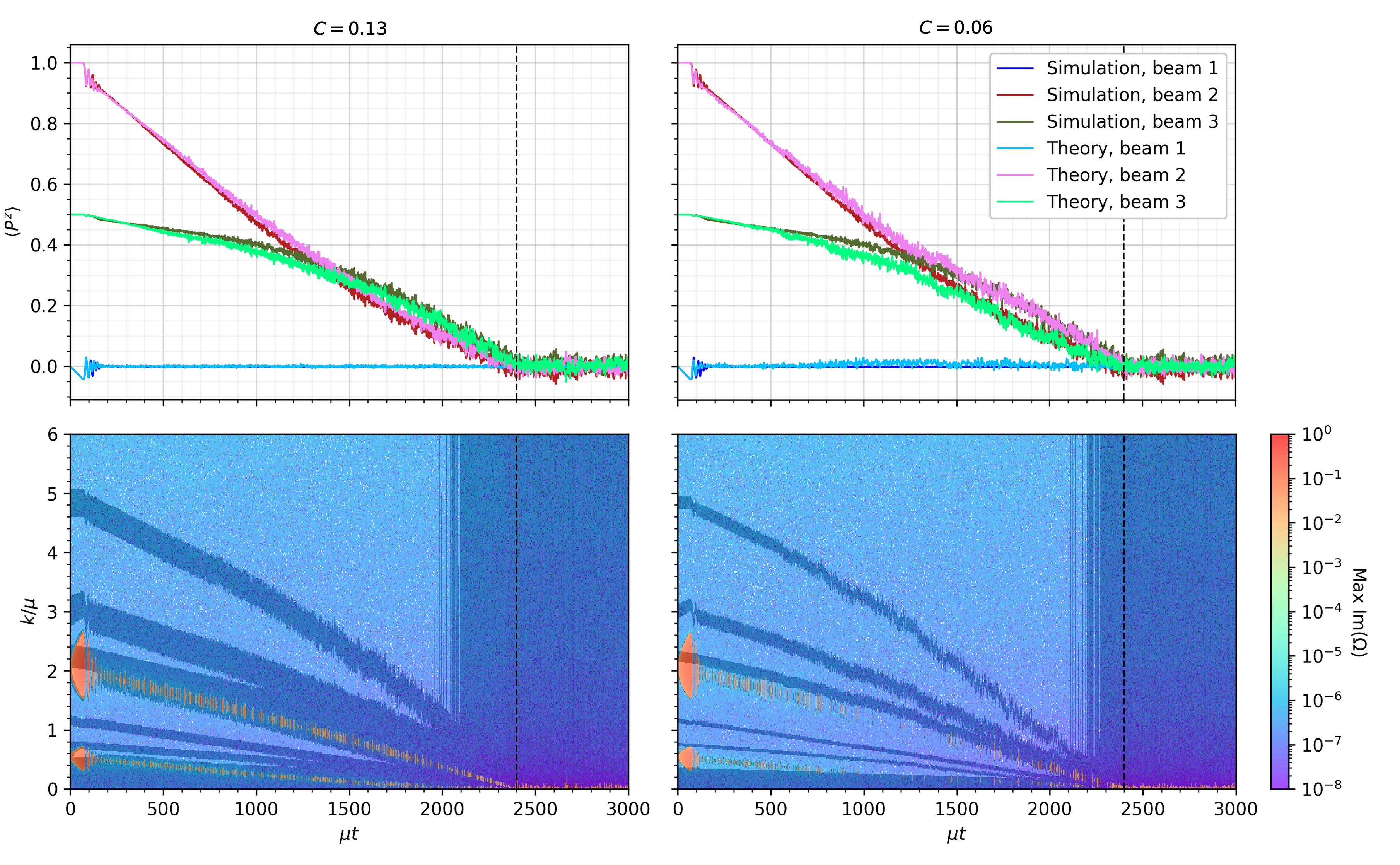}
    
    \caption{Same as Fig.~\ref{fig:QS2beam} but for the three-beam case, with QS thresholds $C = 0.13$ (left) and $C = 0.06$ (right). As in the two-beam case, the QS approximation can be successfully applied to some wave vectors but at no time can it be applied to all of them. The exact evolution is significantly shaped by nonadiabatic transitions amplified by degeneracies and small energy gaps.}
    \label{fig:QS3beam}
\end{figure*}

\begin{figure}
    \centering
    \includegraphics[width=\linewidth]{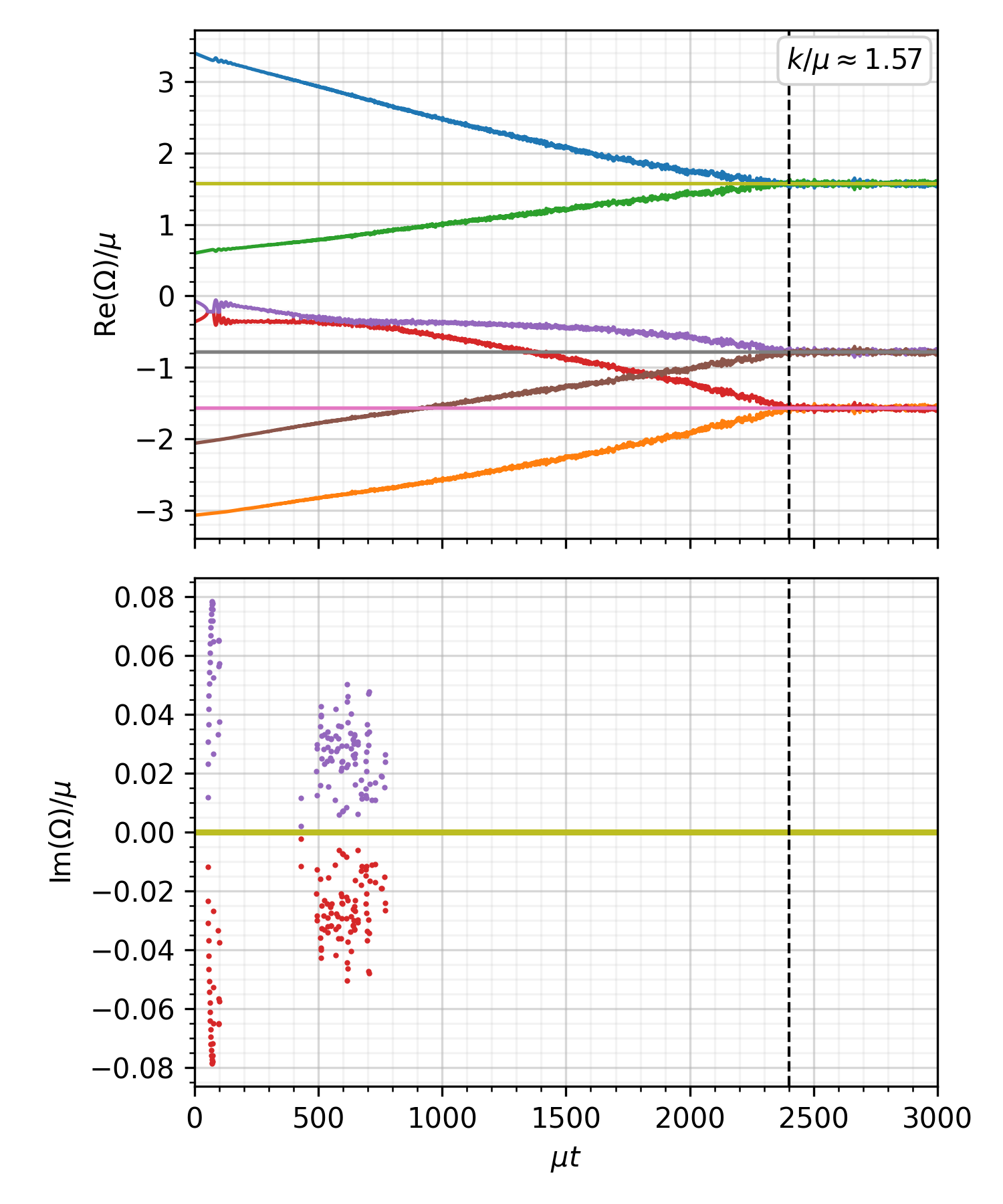}
    
    \caption{Real and imaginary parts of eigenvalues $\Omega / \mu$ at wave vector $k / \mu \approx 1.57$, using data from the three-beam hybrid calculation with $C = 0.13$, as functions of time $\mu t$. Data points are plotted every increment of 1 in $\mu t$.}
    \label{fig:ReImOmega}
\end{figure}

Figures~\ref{fig:QS2beam} and \ref{fig:QS3beam} show the results of the hybrid method for the same two- and three-beam cases analyzed in Sec.~\ref{sec:NL}. For each case, we present hybrid calculations using two different  tolerances for the QS criterion [Eq.~\eqref{eq:QScriterion}]. Focusing first on Fig.~\ref{fig:QS2beam}, we see that the hybrid calculation with QS criterion $C = 0.05$ performs decently well, with numerical agreement comparable to the QL approximation (upper left panel). This agreement occurs despite the fact that the majority of Fourier modes in the simulation are treated using the QS approximation for $\mu t \gtrsim 1600$ (lower left panel). The QS approximation is virtually unused prior to this time due to the prevalence of \mbox{(near-)degeneracies}. With a slightly smaller QS tolerance, $C = 0.03$, the agreement between theory and simulation becomes appreciably worse (upper right panel). This is observed even though the distributions of modes passing and failing the QS criterion are not dramatically different (lower panels).

The conclusions from Fig.~\ref{fig:QS3beam} are similar. At a tolerance of $C = 0.13$, the agreement is quite good (upper left panel). Discrepancies are much more severe at $C = 0.06$, with the $\langle P^z \rangle$ values of beams 2 and 3 failing to cross as seen in the simulation (upper right panel). The $( k/ \mu, \mu t)$ plots for the three-beam case are qualitatively different from those in the two-beam case, with strips cutting through this plane where \mbox{(near-)degeneracies} occur (lower panels).

Figure~\ref{fig:ReImOmega} exemplifies the behavior of the eigenvalues at a particular wave vector during the evolution, here using $k / \mu \approx 1.57$ from the three-beam calculation with $C = 0.13$. See the left panels of Fig.~\ref{fig:QS3beam} for the corresponding plots of $\langle P^z \rangle$ versus $\mu t$ and $\textrm{Max} \, \Omega^{\textrm{I}}$ in the $(k / \mu, \mu t)$ plane. The upper panel of Fig.~\ref{fig:ReImOmega} shows merging and splitting of $\Omega^{\textrm{R}}$ for two modes (the red and purple curves) at early times $0 \lesssim \mu t \lesssim 100$ followed by a later such period spanning $400 \lesssim \mu t \lesssim 800$. These periods, which are bookended by exceptional points if not also punctuated by them, are reflected in the lower panel by nonzero $\Omega^{\textrm{I}}$.

The principal takeaway from our hybrid calculations is that flavor-wave transport theory will either require a tractable way of approximating nonquasistatic level transitions or it will need to be formulated in an alternate way that somehow avoids the problem posed by exceptional points. As Figs.~\ref{fig:QS2beam} through \ref{fig:ReImOmega} demonstrate, spectral degeneracies undermine quasistaticity, leading to worsening results as the QS approximation is applied more aggressively (\textit{i.e.}, as $C$ is made smaller). For avoided level crossings in Hermitian quantum systems, one could use the Landau--Zener formula to approximate finite-rate level transitions. We are not aware of a simple replacement for the Landau--Zener formula that applies to evolution through or near exceptional points.

\section{Rotational symmetry breaking\label{sec:rot}}

Past work on flavor instabilities has highlighted the significance of spontaneous symmetry breaking in phase space \cite{chakraborty2016collective, tamborra2021new}. Here we raise a distinct question: Are new flavor instabilities opened up by spontaneous symmetry breaking in \textit{flavor} space?\footnote{Explicit symmetry breaking due to neutrino masses and matter potentials is addressed within miscidynamics by defining flavor waves on a symmetry-broken mixing equilibrium \cite{johns2023thermodynamics, johns2025local, kost2025once}. Since we focus on FFC in this study, with neutrino masses and matter potentials neglected in the Hamiltonians, symmetry breaking can only occur spontaneously here.}

We noted in Sec.~\ref{sec:spec} that we impose $\langle \bm{P}_{\bm{q}} \rangle \propto \bm{z}$ for all beams in our calculations. Under this approximation, the flavor-wave eigensystems have rotational symmetry about $\bm{z}$. Numerically we have confirmed that in fact the $\bm{n} = \bm{0}$ vectors fluctuate away from the $\bm{z}$-axis, but that these small fluctuations have no meaningful impact on the evolution. We found that our simulation results were virtually unchanged when we forced $\langle \bm{P}_{\bm{q}} \rangle \propto \bm{z}$ by hand. Nonetheless, the existence of these transverse fluctuations prompted us to inspect how they change the flavor-wave eigenanalysis.

Traditionally, linear stability analysis of dense neutrino systems is carried out assuming that $\langle \bm{P}_{\bm{q}} \rangle$ is approximately aligned with $\bm{z}$ and $\bm{P}_{\bm{q},\bm{k}} \cdot \bm{z} = 0$ for all momenta $\bm{q}$ and all wave vectors $\bm{k} \neq \bm{0}$. The linearization is based on the assumption that transverse parts $\bm{P}^T_{\bm{q}}(t, \bm{r})$ are small. In Ref.~\cite{johns2025local} we presented a generalization in which all of these assumptions are relaxed and it is assumed only that $\bm{P}_{\bm{q},\bm{k}}$ at any given $\bm{k}$ is small. This more general framework allows us to search for unstable eigenvalues of the flavor-wave Hamiltonian $\mathcal{H}_{\bm{n}}$ as we vary the transverse parts of the $\langle \bm{P}_{\bm{q}}\rangle$ vectors.

\begin{figure}
    \centering
    \includegraphics[width=\linewidth]{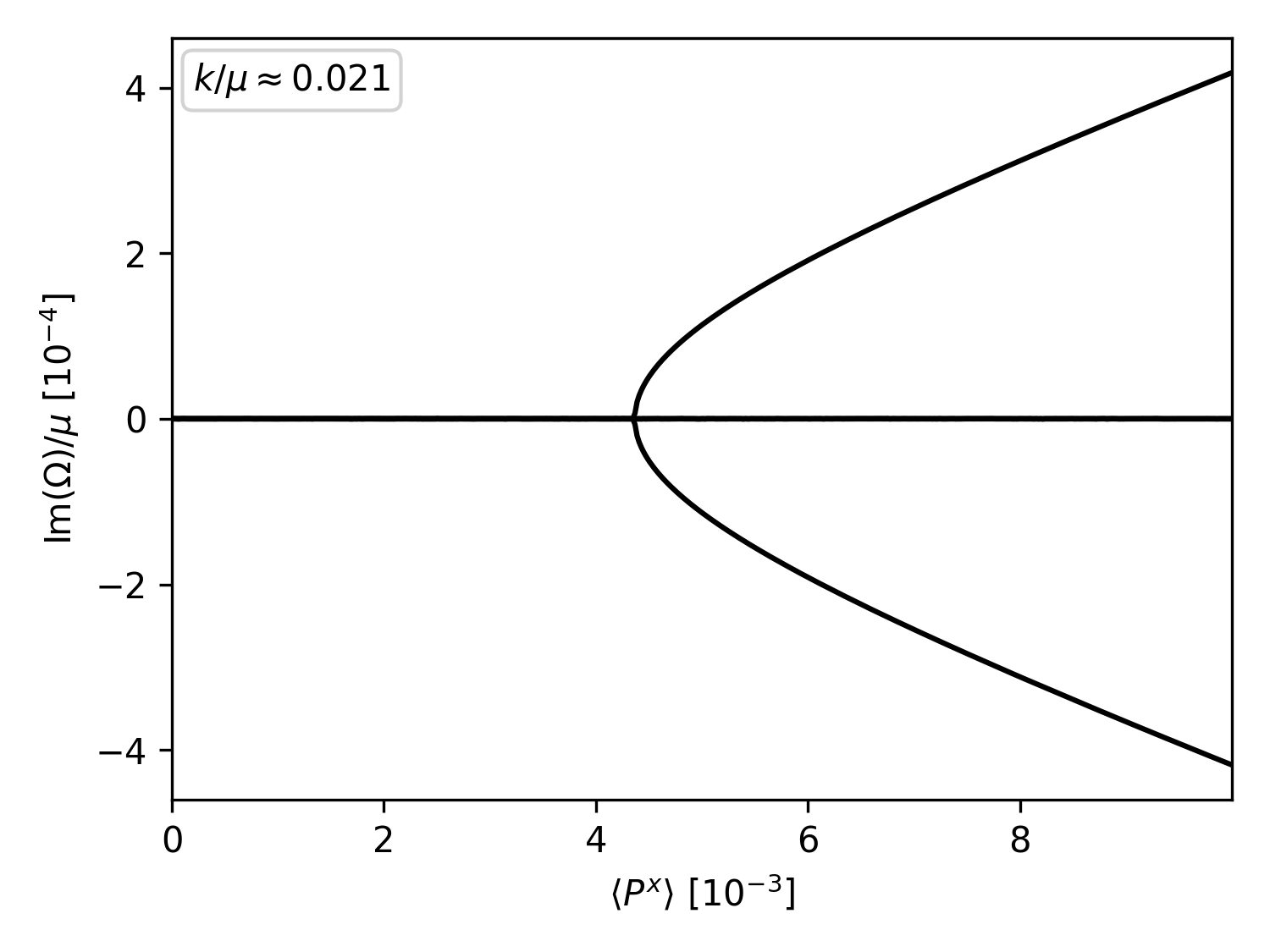}
    \caption{The imaginary parts of the eigenvalues $\Omega / \mu$ for  wave vector $k / \mu \approx 0.021$ in the two-beam setup, as functions of $\expval{P^x}$, where $\expval{\bm P_{L}} = \left(\expval{P^x}, 0, 0.1 \right)$ and $\expval{\bm P_{R}} = \left(\expval{P^x}, 0, 0.5 \right)$. All modes are stable with a standard polarization background that has all mean vectors---in this case $\langle \bm{P}_{\textrm{L}} \rangle$ and $\langle \bm{P}_{\textrm{R}} \rangle$---oriented along $\bm{z}$. Unstable modes emerge once $\langle P^x \rangle$ passes a certain threshold.}
    \label{fig:Im-Omega-vs-Px}
\end{figure}

As a test case, we use a two-beam model with left- and right-going beams whose polarization vectors are $\bm{P}_L$ and $\bm{P}_R$, respectively. In Fig.~\ref{fig:Im-Omega-vs-Px} we plot the growth rates $\Omega^{\textrm{I}}$ as functions of the transverse part $\langle P^x\rangle$ for an illustrative wave vector $k/\mu \approx 0.021$. We set the mean polarizations to be $\expval{\bm P_{L}} = \left(\expval{P^x}, 0, 0.1 \right)$ and $\expval{\bm P_{R}} = \left(\expval{P^x}, 0, 0.5 \right)$. At $\langle P^x\rangle = 0$, all modes are stable ($\Omega^{\textrm{I}} = 0$), as expected based on the absence of an angular crossing. Interestingly, as $\langle P^x\rangle$ is increased, the system goes through a ``phase transition'' as growing and decaying modes suddenly appear in the flavor-wave spectrum at this wave vector. Notably, unstable modes emerge when $\langle P^x\rangle$ is only a small fraction of $\langle P^z_L \rangle$ and $\langle P^z_R \rangle$.

We stress again that we detect no influence in the numerical calculations presented in preceding sections from instabilities associated with spontaneous rotational symmetry breaking. However, we cannot rule out the possibility that they have effects in other models.

\section{Discussion\label{sec:disc}}

In this work we have examined the QL and QS approximations, leaving aside various other approximations that may be useful or even essential for coarse-grained neutrino quantum kinetics. Based on periodic-box calculations, we argue that the QL approximation is of contestable reliability. While the approximation is a substantial simplification over calculating all nonlinear wave--wave interactions, we find it hard to say whether it will be applicable to real astrophysical settings. As for the QS approximation, we find that it may be applied to wave vectors $\bm{k}$ at which all mode splittings are sufficiently large. However, exceptional points are generic in flavor-wave spectra when neutrinos are driven through marginally stable states. Even deep in the stable regime, flavor-wave spectra exhibit small energy gaps. Spectral degeneracies cause a breakdown of the validity of the QS approximation because they amplify mode transitions due to finite rates of driving.

The numerical tests we have presented all involve either two or three discrete momentum bins. It may be that the continuum limit, with an infinite number of momentum bins, behaves fundamentally differently, somehow resolving the problem of nonquasistatic transitions amplified by spectral degeneracies. Either way, it is important to recognize the challenge posed when  momentum discretization is inherited from classical neutrino transport based on Boltzmann methods.

The potential need to go beyond the QL approximation raises the question of whether wave--wave interactions can be approximated in some other way short of neglecting them entirely. We have previously discussed two possible paths forward: flavor-wave kinetics and, even more extreme, flavor-wave thermalization \cite{johns2025local}. Rapidly varying phases generally appear in the nonlinear term $N_{\bm{n}}$. The secular approximation retains only the resonant terms that do not rapidly average out. It leads to flavor-wave kinetics and a collisionlike approximation of wave--wave (and for that matter wave--particle) interactions, isolating the energy-conserving processes with
\begin{equation}
    \Omega^{\textrm{R}}_{\bm{n}-\bm{m},j} + \Omega^{\textrm{R}}_{\bm{m},l} = \Omega^{\textrm{R}}_{\bm{n},i}.
\end{equation}
While the secular approximation eliminates the rapid phase variation, it requires solving the dispersion-relation-dependent resonance condition above to isolate the dominant interactions. In the even more aggressive flavor-wave thermalization hypothesis, wave turbulence is assumed to redistribute flavor waves to a maximum-entropy distribution given certain conserved quantities under wave--wave interactions. We leave numerical tests of the secular and flavor-wave thermalization approximations to future work.

The hybrid calculations presented in Sec.~\ref{sec:spec} assume $\langle \bm{P}_{\bm{q}} \rangle \propto \bm{z}$. We find that this assumption makes the evolution more numerically stable. It is also a reasonable approximation given that the Hamiltonians $\bm{H}_{\bm{q},\bm{0}}$ do not explicitly break flavor-space rotational symmetry around $\bm{z}$. (We dealt in Sec.~\ref{sec:rot} with the question of \textit{spontaneous} breaking of this symmetry.) In scenarios with explicit symmetry breaking, such as when the mass-squared splitting $\delta m^2$ and vacuum mixing angle $\theta$ are nonzero, the configuration $\langle \bm{P}_{\bm{q}} \rangle \propto \bm{z}$ for all momenta is not a fixed point of the $\bm{k} = \bm{0}$ dynamics. Regardless of the flavor-wave populations, we will have
\begin{equation}
    \frac{d}{dt} \langle \bm{{P}}_{\bm{q}} \rangle = \omega_{\bm{q}} \bm{B} \times \langle \bm{P}_{\bm{q}} \rangle + \dots,
\end{equation}
where $\bm{B} = \sin 2\theta \bm{x} - \cos 2\theta \bm{z}$ is the mass basis vector. The term shown on the righthand side prevents $\langle \bm{P}_{\bm{q}} \rangle \propto \bm{z}$ from being a fixed point except under special circumstances. If $\langle \bm{P}_{\bm{q}} \rangle \propto \bm{z}$ initially, it will evolve, and its evolution will in turn feed into the evolution of all other $\langle \bm{P}_{\bm{q'}} \rangle$. In general these dynamics occur on the $\mathcal{O}(\mu^{-1})$ timescale. Flavor-wave eigensystems inherit the rapid variation of the mean polarizations, potentially undermining the QS approximation even far from spectral degeneracies. The idea behind flavor-wave transport, we emphasize again, is to exploit the slow variation of the flavor-wave eigensystems even when the fine-grained evolution may be rapid. The hypothesis of local mixing equilibrium,
\begin{equation}
    \langle\bm{H}_{\bm{q}}\rangle \times \langle\bm{P}_{\bm{q}}\rangle = 0 ~ \textrm{for all} ~ \bm{q} ~~~ (\textrm{mixing equilibrium}),\label{eq:mixeq}
\end{equation}
resolves this issue by ensuring that mean polarizations evolve only due to astrophysical driving and flavor-wave viscosity \cite{johns2023thermodynamics, johns2024subgrid, johns2025local}. We formulated flavor-wave transport in the context of miscidynamics, assuming that the mean polarizations are in mixing equilibrium but not necessarily along $\bm{z}$ \cite{johns2025local}. The numerical calculations of Ref.~\cite{kost2025once} do indeed show the emergence and tracking of instantaneous mixing equilibria that are not polarized strictly along $\bm{z}$ (see the precedent for such equilibria in Refs.~\cite{raffelt2007self2, raffelt2007adiabaticity}). Further testing remains to be done of the local-equilibrium hypothesis and of flavor-wave transport on other backgrounds besides those strictly polarized along $\bm{z}$.

The numerical test cases in this paper all involve spatially periodic systems. This has been a common approach to gaining insights into neutrino flavor evolution. In Ref.~\cite{johns2025local} we introduced a way to transport flavor waves, which are defined locally, between adjacent regions (see Ref.~\cite{fiorillo2026flavomon} for further development of this proposal). The grid-level propagation of flavor waves may be treated using the ray approximation in which the $i$th eigenmode at wave vector $\bm{k}$ travels with group velocity $\bm{v}_i^g$ and force $\bm{F}_i$
\begin{equation}
    \frac{d\bm{r}}{dt} = \bm{v}_i^g \equiv \frac{\partial \Omega_i^{\textrm{R}}}{\partial \bm{k}}, ~~~ \frac{d\bm{k}}{dt} = \bm{F}_i \equiv -\frac{\partial \Omega_i^{\textrm{R}}}{\partial \bm{k}}.
\end{equation}
We leave numerical tests of this approximation, and an examination of spectral degeneracies encountered during propagation through the macroscopic environment, for future work as well.

Flavor-wave transport is an appealing approach to approximating small-scale neutrino flavor dynamics. It was recently shown to have definite potential for correctly reproducing some behaviors of the full QKEs \cite{fiorillo2026quasi}. When combined with miscidynamics, it completely coarse-grains over the small-scale dynamics of the QKEs. However, we have shown here that it encounters challenges in scenarios where neutrinos experience slow astrophysical driving. While the QL approximation is dispensable, the QS approximation is essential to flavor-wave transport as currently formulated. It will be imperative to find a way to approximate or bypass the nonquasistatic evolution associated with spectral degeneracies.

\begin{acknowledgments}
We thank Jiabao Liu and Hiroki Nagakura for helpful conversations. A.~K. and H.~D. are supported by the US DOE NP grant No. DE-SC0017803 at UNM. L.~J. is supported by the US Department of Energy and Los Alamos National Laboratory under contract 89233218CNA000001 and by a Feynman Fellowship through LANL LDRD project No. 20230788PRD1.
\end{acknowledgments}

\bibliography{refs}

\end{document}